\documentclass[aps,prb,preprint,notitlepage]{revtex4-2}
\usepackage{amsmath}
\usepackage{graphicx}
\usepackage{amssymb}

\usepackage{gt17}
\usepackage{color}
\usepackage{graphicx}% Include files
\usepackage{bm}% bold math
\usepackage{amsmath}
\usepackage{ulem} % sout
\begin{document}
%%%%%%%%%%%%%%%%%%%%%%%%%%
%%%%%%%%%%%%%%%%%%%%%%%%%%%%%%%%%%%%%
\title{
Vorticity-induced anomalous Hall effect in electron fluid
}
\author{Hiroshi Funaki$^{1}$}
%\email{gen.tatara@riken.jp}
\author{Riki Toshio$^{2}$}
\author{Gen Tatara$^{1,3}$}
\affiliation{$^{1}$ RIKEN Center for Emergent Matter Science (CEMS),
2-1 Hirosawa, Wako, Saitama, 351-0198 Japan}
\affiliation{$^{2}$ Department of Physics, Kyoto University, Kyoto 606-8502, Japan}
\affiliation{$^{3}$ RIKEN Cluster for Pioneering Research (CPR),
2-1 Hirosawa, Wako, Saitama, 351-0198 Japan}
\date{\today}

\begin{abstract}
We develop a hydrodynamic theory for an electron system exhibiting the anomalous Hall effect, and show that an additional anomalous Hall effect is induced by a vorticity generated near boundaries.
We calculate the momentum flux and force proportional to the electric field using linear response theory. The hydrodynamic equation is obtained by replacing the local electric field with the electric current, focusing on a scale that is sufficiently larger than the mean free path.
It is demonstrated that there is a coupling between a vorticity of an electric current and a magnetization which generates a pressure from non-uniform vorticity.
Taking into account Hall viscosity and relaxation forces, a non-uniform flow near a boundary and an additional Hall force are calculated.
The additional anomalous Hall force is opposite to conventional anomalous Hall force, resulting in a sign reversal in thin systems.
\end{abstract}

\maketitle
%%%%%%%%%%%%%%%%%%%%%%%%%%%%
\newcommand{\chiv}{{\bm \chi}}
\newcommand{\fv}{{\bm f}}
\newcommand{\omegav}{{\bm \omega}}
\newcommand{\sigmae}{\sigma_{\rm e}}
\renewcommand{\vimp}{u_{\rm i}}
%\newcommand{\red}[1]{\textcolor{red}{#1}}
%\newcommand{\out}[1]{\textcolor{red}{\sout{\textcolor{black}{#1}}}}
%%%%%%%%%%%%%%%%%%%%%%%%%
\section{Introduction}
%%%%%%%%%%%%%%%%%%%%%%%%%%
%\subsection{Spin-vorticity coupling}
%%%%%%%%%%%%%%%%%%%%%%%%%%

Spin, a form an angular momentum, leads to various exotic phenomena  different from those of charge.
A typical example is the spin-rotational coupling, where the angular momentum of mechanical rotation couples directly to spin, inducing classical Einstein-de Haas effect.
Spin-rotation coupling was derived from the Dirac equation in a rotating frame in Ref. \cite{MatsuoPRL11}.
A spin-vorticity coupling to the vorticity of the electron flow, $\omegav=\nabla\times\jv_{\rm e}$, where $\jv_{\rm e}$ is the electron current density, was derived and spin current generation was discussed recently \cite{MatsuoHydro17}.
Spin vorticity coupling is natural from the view point of the Faraday's law, $\dot{\Bv}=-\nabla\times \Ev$ between an electric and magnetic fields, $\Ev$ and $\Bv$, respectively.
In fact, the right-hand side is proportional to the vorticity of electric current, as local electric current density is related in the linear response regime to the electric field as $\jv_{\rm e}=\sigmae\Ev$ where $\sigmae$ is the electric conductivity and spin density is induced by a magnetic field.

Spin Hall effect, where spin density \cite{DyakonovPhysLett71}
and current \cite{Hirsch99} are induced by an applied electric field, turned out recently to be interpreted as due to the spin-vorticity coupling \cite{TataraSH18}. It was shown there that the induced spin density $\sv$ in the ballistic regime is written as
$\sv=\lambda_{\rm sh}(\nabla\times \jv_{\rm e})$, where $\lambda_{\rm sh}$ is a constant arising from the spin-orbit interaction, indicating an effective spin-vorticity coupling of the form $\sv\cdot\omegav$.
(In the diffusive case, the coupling exists but becomes nonlocal due to diffusion.)
As is expected from the symmetry and the Faraday's law, $\lambda_{\rm sh}$ is proportional to the electron relaxation time $\tau$, meaning that a relaxation is essential for the development of spin density.
The above studies indicate that spin-vorticity coupling, arising as a natural consequence of the spin-orbit interaction, is a fundamental coupling for various spintronics effects.

Spin-vorticity coupling suggests that there is also a magnetization-vorticity coupling inducing a rotational motion due to a magnetization in ferromagnets.
The aim of the present paper is to demonstrate that such a coupling indeed emerges as a result of an anomalous Hall effect.
Vorticity of electrons in solids has not been discussed in the context of conventional electron transport properties focusing on the spatially averaged responses.
In mesoscopic systems and interface/surface transports, in contrast, vorticity of flow in the meso or macroscopic size would be crucially important.
Hydrodynamic description, which integrates out microscopic features, become a powerful tool to take account of such vortical effects.
In fact, recent intense studies of {\it electron hydrodynamics}, which is quickly growing into a mature field of condensed matter physics today~\cite{Lucas18,Polini20}, has predicted and demonstrated various unconventional transport phenomena driven by vorticity or velocity gradient, including negative local resistance~\cite{Torre15, Bandurin16,Levitov16,Bandurin18}, anomalous viscous magnetotransport~\cite{Alekseev16,Scaffidi17,Pellegrino17,Moll16,Gooth18,Berdyugin19}, spin hydrodynamic generation~\cite{Matsuo20}, generalized vortical effect~\cite{Toshio20,Shitade20}, and chiral angular momentum generation~\cite{Funaki20}.
In this work, we choose a hydrodynamic approach for the analysis of the Ohmic fluid in anomalous Hall (AH) systems and evaluate the momentum flux density based on a microscopic linear response theory, as done in Ref. \cite{Funaki20}.

%%%%%%%%%%%%%%%%%%%%%%%%%%
\section{Anomalous Hall electron fluid}
%%%%%%%%%%%%%%%%%%%%%%%%%%
Electrons in solids are dense and are regarded as continuum medium or a fluid.
Macroscopic or mesoscopic transport properties of electron fluid is characterized by the forces acting on the fluid.
Besides external forces such as the one due to the electric field and the Lorentz force, there are viscosity force and relaxation force induced by internal interactions and scatterings.
Viscosity force arises from inhomogeneity of fluid velocity.
In systems with high symmetry, viscosity tensor is symmetric, while  antisymmetric component of the viscosity (rotational viscosity) arises when inversion symmetry is broken \cite{MatsuoHydro17,Doornenbal19}.
In chiral systems a bulk rotational force arises from linear velocity, resulting in a chiral angular momentum generation \cite{Toshio20,Funaki20}.
Viscosity is larger for longer relaxation time (weaker relaxation), as the relaxation effects cut off interaction effects that lead to viscosity.
The relaxation force is opposite to the fluid velocity and is written as
$\fv_{\rm r}=-m\jv/\tau$, where $m$ is the electron mass, $\tau$ is the total relaxation time and
$\jv$ is the current density without the electron charge $e$ ($e\jv=\jv_{\rm e}$).
Depending on the relaxation time, therefore, electron fluids are classified into two regimes, a viscous fluid and ohmic fluid, where viscosity and relaxation  dominates, respectively.
Most metals are in the ohmic regime, while viscous fluids have realized recently in extremely clean systems such as graphene \cite{Polini20, Berdyugin19, Bandurin16, Crossno16, Kumar17, Sulpizio19}, GaAs quantum wells~\cite{Molenkamp94,Jong95}, 2D monovalent layered metal PdCoO${}_2$~\cite{Moll16}, and various semimetallic materials including WP${}_2$~\cite{Gooth18}, WTe${}_2$~\cite{Uri20}, MoP~\cite{Kumar19}, Sb~\cite{Jaoui20}, and ZrTe${}_5$~\cite{Chang20}.

Here we study an electron fluid showing anomalous Hall effect due to a uniform magnetization $\Mv$.
Considering the ohmic regime, we take account of a spin-orbit interaction arising from impurities, on the same footing as the studies of anomalous Hall conductivity \cite{Dugaev01}.
Calculating the momentum flux density within the linear response theory, we show that the anomalous Hall liquid has a magnetization-vorticity coupling, $\Mv\cdot\omegav$, besides an anomalous viscosity  argued previously \cite{Scaffidi17}.
The magnetization-vorticity  coupling is a spin-polarized counterpart of the spin-vorticity coupling.
In terms of the force density, the contribution reads
$\tilde{\lambda}\nabla(\hat{\Mv}\cdot\omegav)$ with a coefficient $\tilde{\lambda}$.
In the case of a thin film ferromagnet with an in-plane magnetization with an applied electric field, this coupling induces a voltage perpendicular to the magnetization and the applied electric field.
The direction of the output voltage is opposite to the conventional bulk anomalous Hall effect.
The reduction of Hall effect due to a Hall viscosity was reported in Ref.~\cite{Scaffidi17}.
We also identify an anomalous Hall force  $f_{\rm ah}(\hat{\Mv}\times\jv)$ ($f_{\rm ah}$ is a coefficient and $\hat{\Mv}\equiv \Mv/M$), which is analogous to the Lorentz force.
The total force density acting on the anomalous Hall fluid with uniform and steady flow is therefore
$\fv=en\Ev-\frac{m}{\tau}\jv+f_{\rm ah}(\hat{\Mv}\times \jv)$.
When an electric field $\Ev$ is perpendicular to the magnetization,
the steady flow realized is
$e \jv=\sigmae E+\sigma_{\rm ah}(\hat{\Mv}\times \Ev)$ with $\sigma_{\rm ah}/\sigmae=\frac{\tau}{m}f_{\rm ah}$ to the lowest order in the spin-orbit interaction.

The magnetization-vorticity coupling found here is a potential for the electron, and thus does not generate angular momentum. In fact, the  magnetization-vorticity coupling induces a pressure $\fv_{\omegav}=\tilde{\lambda} \nabla(\omegav\cdot\Mv)$, and its contribution to the orbital angular momentum vanishes, as
$\int d^3r (\rv\times \nabla(\omegav\cdot\Mv))=0$ if we use integral by parts.
This is because the coupling is between two angular momenta, and thus linear velocity does not generate angular momentum.
In contrast, in chiral systems, an angular momentum couples to a linear momentum, resulting in a linear angular momentum generation \cite{Toshio20,Funaki20}.

%%%%%%%%%%%%%%%%%%%%%%%%%%
\section{Formalism}
%%%%%%%%%%%%%%%%%%%%%%%%%%
The model we consider is a conduction electron with a spin polarization due to a uniform localized spin (magnetization) $\Sv$ and spin-orbit interaction.
The Hamiltonian for the electron is $H\equiv H_K+H_M+H_{\rm so}+H_{\rm i}$, where
\begin{align}
 H_K &=\intr c^\dagger \frac{-\nabla^2}{2m} c
\end{align}
is the kinetic part, and
\begin{align}
H_{M}\equiv -\sum_{\kv}  c^\dagger_{\kv} (\Mv\cdot\sigmav) c_{\kv}
\end{align}
is the exchange interaction to the magnetization, where $\Mv\equiv J\Sv$ with $J$ being  a coupling constant.
The spin-orbit interaction is the one arising from random impurities, represented by a Hamiltonian
\begin{align}
 H_{\rm so}
 &= \lambda_{\rm so} \int d^3r  c^\dagger(\rv) [(\nabla \vimp(\rv) \times \hat{\pv}) \cdot\sigmav] c(\rv)  \nnr
&= i\lambda_{\rm so} \sum_{\kv\kv'} u_{\kv'-\kv} (\kv'\times\kv) \cdot c^\dagger_{\kv'} \sigmav c_{\kv}
\end{align}
where $\vimp(\rv)=\vimp\sum_{\Rv_i}\delta(\rv-\Rv_i)$ is an impurity potential, where $\vimp
$ is the strength of the impurity potential and $\Rv_i$ is the position of  $i$-th impurity.
$u_{\kv'-\kv}=\vimp\sum_{\Rv_i}e^{-i(\kv'-\kv)\cdot\Rv_i}$ is the Fourier transform of the impurity potential.
The spin-orbit interaction leads to a correction to the electric velocity operator (anomalous velocity),
\begin{align}
 \delta \vv(\rv) &=-i\lambda_{\rm so}  \sum_{\kv\kv'\qv} u_\qv e^{i(\kv-\kv'+\qv)\cdot\rv} c^\dagger_{\kv'} (\qv\times\sigmav) c_{\kv}
\end{align}
Considering diffusive (Ohmic) regime, impurity scattering Hamiltonian,
\begin{align}
 H_{\rm i} &= \sum_{\kv\kv'} u_{\kv'-\kv} c^\dagger_{\kv'} c_{\kv}
\end{align}
is included.
The elastic life time arising form the impurity scattering, $\taue$, is given by $\taue^{-1}=2\pi\nu \vimp^2 n_{\rm i}$, where $\nu$ and $n_{\rm i}$ are density of states and impurity concentration, respectively.
We treat it as spin-independent to simplify the calculation.

We study the effects of the anomalous Hall effect on the hydrodynamic behaviors of the electron taking account of both the spin-orbit interaction and an applied electric field to the linear order.
A hydrodynamic equation is derived by calculating the equation of motion for the momentum density, $\pv\equiv \average{c^\dagger \hat{\pv}c}$ ($\hat{\pv}\equiv -i\nabla$ is the momentum operator) by use of the Heisenberg equation of motion,
$\dot{\pv}=i[H,c^\dagger \hat{\pv}c]$.
The commutator with the kinetic part is
\begin{align}
  i[H_K,c^\dagger \hat{p}_i c] &= -\nabla_j \pi^0_{ij}
\end{align}
where the momentum flux density contribution is
\begin{align}
 \pi^0_{ij}(\rv,t) &= -i \frac{1}{m}\tr[\hat{p}_i \hat{p}_j G^<(\rv,t,\rv,t)]
\end{align}
with $G^<(\rv,t,\rv',t')=i\average{c^\dagger(\rv',t')c(\rv,t)}$ being the lesser Green's function and $\tr$ is the summation over spin.
The spin-orbit contribution is
\begin{align}
  i[H_{\rm so} ,c^\dagger \hat{p}_i c] &= -\nabla_j \pi^{\rm so}_{ij} +f^{\rm so}_i
\end{align}
where the momentum flux density contribution from the spin-orbit interaction is
\begin{align}
 \pi^{\rm so}_{ij}(\rv,t)
 %&= - i\frac{\lambda_{\rm so}}{2}\epsilon_{jkl} (\nabla_j v(\rv))
 % [c^\dagger \stackrel{\leftrightarrow}{\sigma_l} \nabla_i c] \nnr
  &= \frac{\lambda_{\rm so}}{2}\epsilon_{jkl} (\nabla_{k} \vimp(\rv))
 (\nabla_i^{\rv}-\nabla_i^{\rv'})\tr[\sigma_l G^<(\rv,t,\rv',t)]|_{\rv'\ra\rv}
\end{align}
The term $f^{\rm so}_i$ is a contribution not written as a divergence, which is
\begin{align}
f^{\rm so}_i(\rv,t)
%&=  \frac{\lambda_{\rm so}}{2}\epsilon_{jkl}
%(\nabla_i \nabla_j \vimp(\rv))   (\nabla_k^{\rv}-\nabla_k^{\rv'})\tr[{\sigma_l} G^<(\rv,t,\rv',t)]|_{\rv'\ra\rv}
&= - \frac{\lambda_{\rm so}}{2}\epsilon_{jkl}
(\nabla_i \nabla_j \vimp(\rv))   (\nabla_k^{\rv}-\nabla_k^{\rv'})\tr[{\sigma_l} G^<(\rv,t,\rv',t)]|_{\rv'\ra\rv}
\end{align}
The impurity contribution leads to a force
\begin{align}
f^{\rm i}_i  &= -i (\nabla_i  \vimp) \tr[G^<(\rv,t,\rv,t)]
\end{align}
The hydrodynamic equation of the present system is therefore
\begin{align}
 \dot{p}_i &= -\nabla_j \pi_{ij}+f_i^{\rm so}+f_i^{\rm i}
 \label{pdoteq}
\end{align}
where the momentum flux density is
\begin{align}
\pi_{ij} &=\pi_{ij}^0+\pi_{ij}^{\rm so} \nnr
&=  -i\tr[\hat{p}_i \hat{v}_j G^<(\rv,t,\rv,t)]
 \end{align}
where $\hat{v}\equiv \frac{\hat{\pv}}{m}+\hat{\delta \vv}$ is the total velocity operator and
$\pi_{ij}^0$ is the normal contribution without spin-orbit interaction.

%%%%%%%%%%%%%%%%%%%%%%%%%%%%%%%%%%%%%%%%%%%%%%%%%%%%%%%
\begin{figure}
 \includegraphics[width=0.3\hsize]{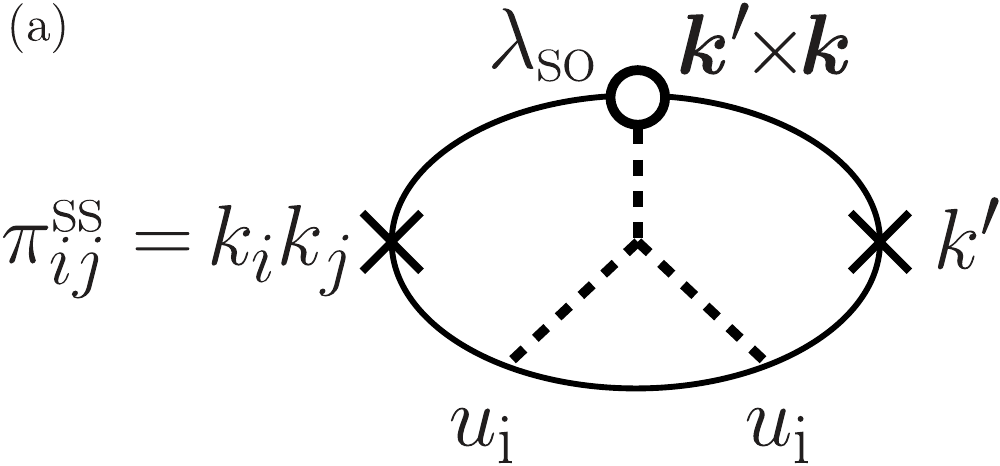}
 \includegraphics[width=0.3\hsize]{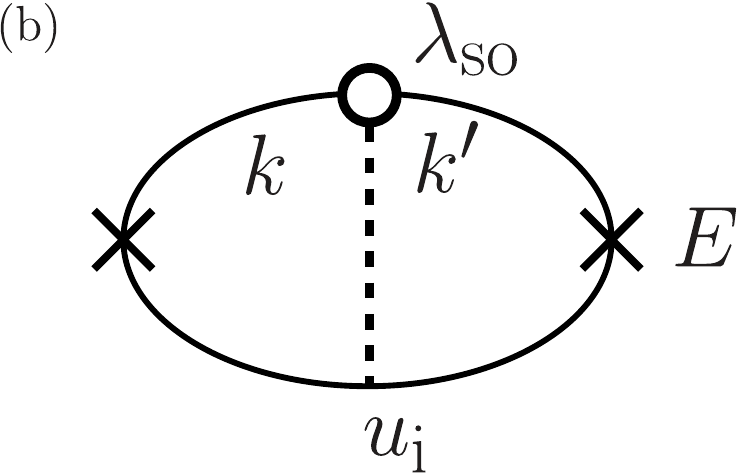}
 \includegraphics[width=0.3\hsize]{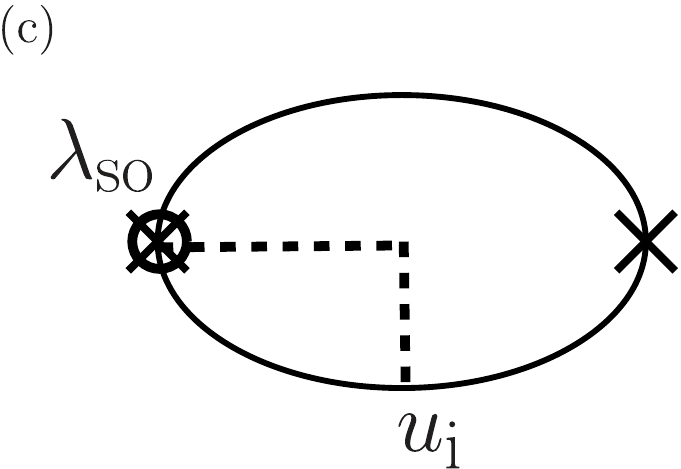}
 \caption{
Feynman diagrams for the dominant contribution to the momentum flux density at the linear order in the spin-orbit interaction and linear response to the applied field, denoted by $\times$ at the right end. Solid lines represent electron Green's functions, where upper and lower lines denote retarded and advanced Green's functions, respectively, and $\kv$ and $\kv'$ are the electron wave vectors. Complex conjugate processes (turned upside down) are also taken into account.
(a): The dominant contribution (skew-scattering).  The left vertex with $k_ik_j$ represents the vertex for the momentum flux density, and  a vertex with $\lambda$ denotes the spin-orbit interaction.
(b): The contribution that vanish when a complex conjugate process (upside-down) is summed.
(c): The contribution arising from the anomalous velocity $\delta v$ denoted by the left vertex, called the side-jump contribution. It is smaller than the one in (a) by a factor of $(\ef\taue)^{-1/2}$.
\label{FIGpvdiag}}
\end{figure}
%%%%%%%%%%%%%%%%%%%%%%%%%%%%%%%%%%%%%%%%%%%%%%%%%%%%%%%%

%%%%%%%%%%%%%%%%%%%%%%%%%%%%%%%%%%%%%%%%%%%%%%%%%%%%%%%%
\section{Derivation of hydrodynamic equation}
%%%%%%%%%%%%%%%%%%%%%%%%%%%%%%%%%%%%%%%%%%%%%%%%%%%%%%%%

The momentum flux density $ \pi_{ij}$ is calculated in the presence of a driving field, an applied static electric field $\Ev$.
In the linear response theory, we have $ \pi_{ij}=\pi_{ijk}eE_{k}$, where $ \pi_{ijk}$ is the correlation function of $\hat{p}_i\hat{v}_j$ and $\hat{v}_k$.
The calculation of the response function is parallel to that of the anomalous Hall conductivity $\sigma_{xy}$ in Ref. \cite{Dugaev01}.
There are two processes, one arising from the normal velocity $\frac{\hat{\pv}}{m}$, the contribution historically called skew scattering contribution, and the other arising from the anomalous velocity $\hat{\delta\vv}$, called the side-jump contribution.
The dominant  contribution at dilute impurity concentration turns out to be the skew scattering one containing impurity scattering to the second order besides the spin-orbit interaction \cite{Dugaev01}, diagrammatically depicted in Fig. \ref{FIGpvdiag}(a).
The processes with less impurity shown in Fig.  \ref{FIGpvdiag}(b) vanish as the factor of $\kv'\times\kv$ in the spin-orbit interaction changes sign for the two conjugate processes.
The skew scattering contribution  is thus (ss denotes skew-scattering, and $V$ is the system volume)
\begin{align}
 \pi_{ijk}^{\rm ss}(\qv)&=
 \frac{1}{( 2\pi V )^3}
 \frac{i \lambda_{\rm so} \vimp^{3} n_{\rm i}}{m^2}
 \sum_{\kv\kv'\kv''}
 k_i k_j k'_k(\kv'\times\kv)_l
 \tr[\sigma_l G_{\kv+\frac{\qv}{2}}^\ret G_{\kv'+\frac{\qv}{2}}^\ret
 G_{\kv'-\frac{\qv}{2}}^\adv G_{\kv-\frac{\qv}{2}}^\adv (G_{\kv''}^\ret - G_{\kv''}^\adv)]
% \tr[G_{\kv+\frac{\qv}{2}}^\ret \sigma_l G_{\kv'+\frac{\qv}{2}}^\ret
% G_{\kv'-\frac{\qv}{2}}^\adv G_{\kv''}^\adv G_{\kv-\frac{\qv}{2}}^\adv]
 \label{pidef2}
\end{align}
where $G_{\kv}^\ret \equiv [-\frac{k^2}{2m}+\Mv\cdot\sigmav+\frac{i}{2\taue}]^{-1}$ is  the retarded Green's function with elastic lifetime $\taue$ and $G_{\kv}^\adv \equiv (G_{\kv}^\ret)^* $.
We choose $\Mv$ along the $z$ axis and calculate the response function to the lowest order in the external wave vector $\qv$.
Summation over the wave vectors are carried out as
\begin{align}
 \frac{1}{V} \sum_{\kv}G_{\kv}^\adv &=i\pi\nu , \;\;\;
 \frac{1}{V} \sum_{\kv}k_ik_jG_{\kv}^\ret G_{\kv}^\adv = \frac{2\pi}{3}\nu \kf^2\taue \delta_{ij}\nnr
 \frac{1}{V} \sum_{\kv}k_ik_jk_k G_{\kv+\frac{\qv}{2}}^\ret G_{\kv-\frac{\qv}{2}}^\adv
   &=
 - i \frac{2\pi}{15m}\nu \kf^{4}\taue^{2} (\delta_{ij}q_{k}+\delta_{ik}q_{j}+\delta_{jk}q_i)
\end{align}
where $\nu$ and $k_{F}$ are the spin-dependent density of states and Fermi wave vector, respectively.
The result is
\begin{align}
 \pi_{ijk}^{\rm ss}(\qv)
 &= i \lambda (\delta_{ij}\epsilon_{lkz}q_l +\epsilon_{ikz}q_j +\epsilon_{jkz}q_i)
 \label{pijk2}
\end{align}
where the coefficient is
($\pm$ denote the direction of spin along the z axis)
\begin{align}
 \lambda \equiv \frac{1}{45}\frac{\lambda_{\rm so} \vimp^3 n_{\rm i}}{m^3}
 \sum_\pm (\pm)(\nu_\pm )^{3}(k_{{\rm F}\pm} )^6 \taue^{3}
\end{align}
The contribution vanishes if there is no spin polarization ($M=0$).

The contribution arising from the anomalous velocity $\delta \vv$ (the side-jump contribution, depicted in Fig. \ref{FIGpvdiag}(c))  is
\begin{align}
 \pi_{ijk}^{\rm sj}(\qv)&=
 \frac{-i}{(2 \pi V)^2} \frac{\lambda_{\rm so} \vimp^{2} n_{\rm i}}{2m}
 \Re \sum_{\kv\kv'}
 (k'+k+q)_i \epsilon_{jlz}(k'-k)_l \lt(k+\frac{q}{2}\rt)_k
 \tr[\sigma_z G_{\kv'}^\ret G_{\kv}^\ret G_{\kv+\qv}^\adv]
 \label{sjdef}
\end{align}
which reduces to
\begin{align}
% \pi_{ijk}^{\rm sj}(\qv)&=  i\eta^{\rm sj}(q_i \epsilon_{jkz}+\delta_{ik}\epsilon_{jlz}q_l-\epsilon_{ijz}q_k)
%\\
 \pi_{ijk}^{\rm sj}(\qv)&=  i\eta^{\rm sj}
  \lt(-\frac{2}{3} \epsilon_{jiz} q_k +\delta_{ik}\epsilon_{jlz}q_l +\epsilon_{jkz}q_i \rt)
 \label{sj1}
\end{align}
where
\begin{align}
 \eta^{\rm sj} \equiv \frac{\lambda_{\rm so}\vimp^{2} n_{\rm i}}{30 m^2} \sum_\pm(\pm)(\nu_\pm)^2 k_{{\rm F}\pm}^4\taue^2
\end{align}
The side-jump contribution has an asymmetric contribution with respect to $i$ and $j$, although the magnitude  proportional to $\vimp^{2}\taue^2$ is smaller than the skew scattering one by a factor of $(\ef\taue)^{-1/2}$ (noting that $\vimp\propto \taue^{-1/2}$).

The  contribution to the spin-orbit induced momentum flux density
 $\pi_{ij}^{\rm so} =\pi_{ij}^{\rm ss}+\pi_{ij}^{\rm sj}$ is therefore
\begin{align}
\pi_{ij}^{\rm so}(\rv)
 &= e\lambda \lt[ \delta_{ij} (\nabla\times\Ev) \cdot \hat{\Mv}
 + \nabla_j(\Ev \times \hat{\Mv})_i \rt]
 +(e\lambda +e\eta^{\rm sj}) \nabla_i(\Ev \times \hat{\Mv})_j
\nnr
 &+
 e\eta^{\rm sj} \lt[
 \frac{2}{3} \epsilon_{ijz} \hat{\Mv} (\nabla \cdot \Ev)
 - (\hat{\Mv} \times \nabla)_j E_i \rt]
 \label{pij1}
\end{align}
where $\hat{\Mv}\equiv \Mv/|\Mv|$.

%%%%%%%%%%%%%%%%%%%%%%%%%%%%%%%%%%%%%%%%%%%%%%%%%%%%%%%
\begin{figure}
 \includegraphics[width=0.3\hsize]{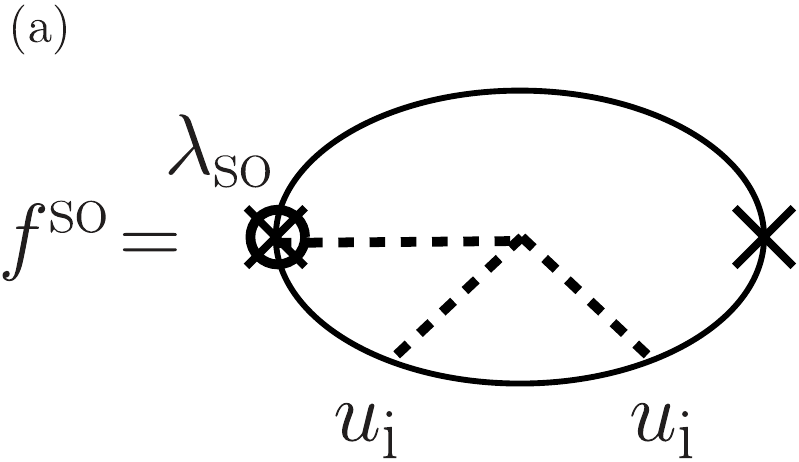}
 \includegraphics[width=0.3\hsize]{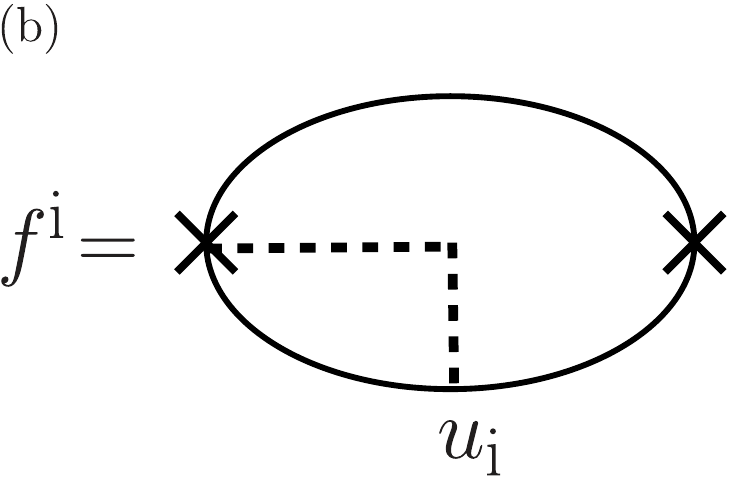}
 \caption{
Feynman diagrams for force due to (a) the spin-orbit interaction and (b) the impurities at the linear order at the linear response to the applied field.
\label{FIGfdiag}}
\end{figure}
%%%%%%%%%%%%%%%%%%%%%%%%%%%%%%%%%%%%%%%%%%%%%%%%%%%%%%%%
Force densities are similarly calculated.
The linear response contribution to a uniform component of the spin-orbit induced force density is $f_{i}^{\rm so} = f_{im}^{\rm so} E_m$, where
(diagrammatically shown in Fig. \ref{FIGfdiag})
\begin{align}
%f_{im}^{\rm so} &\equiv \frac{1}{2\pi}\frac{\lambda_{\rm so}\vimp^2}{m}\sum_{\kv\kv'\kv''} \epsilon_{jkl}
%(k-k')_i k_j k'_k  k_m \tr[\sigma_l (G_{\kv'}^\ret G_{\kv''}^\ret + G_{\kv'}^\adv G_{\kv''}^\adv) G_{\kv}^\ret G_{\kv}^\adv]
f_{im}^{\rm so} &\equiv
\frac{-1}{(2\pi V)^3}\frac{\lambda_{\rm so}\vimp^{3} n_{\rm i}}{m}
\sum_{\kv\kv'\kv''} \epsilon_{jkl}
 (k'-k)_i (k'-k)_j (k'+k)_k k'_m
 \tr[\sigma_l (G_{\kv}^\ret G_{\kv''}^\ret + G_{\kv}^\adv G_{\kv''}^\adv) G_{\kv'}^\ret G_{\kv'}^\adv]
\end{align}
After summation over the wave vectors,
we obtain $f_{im}^{\rm so}=\epsilon_{imz}f_{\rm ah}$, with  a coefficient
\begin{align}
f_{\rm ah} &= \frac{\lambda_{\rm so}\vimp^3 n_{\rm i}}{18m} \sum_{\pm}(\pm)\nu_\pm^{3} k_{{\rm F}\pm}^4 \taue
\end{align}
The force thus is the anomalous Hall force
\begin{align}
 \fv^{\rm so} &= ef_{\rm ah} (\Ev \times \hat{\Mv})
 \label{fsoresult}
\end{align}
The relaxation force due to the impurities turns out to be $\fv^{\rm i}= -en\Ev$, as was argued in Refs. \cite{Gurzhi63,Funaki20}.

Taking account of the normal viscosity $\pi_{ij}^{0}$ and relaxation force, Eqs. (\ref{pdoteq})(\ref{pij1})(\ref{fsoresult}) describes the fluid as a response to the driving field $\Ev$.
Conventional hydrodynamic equation, a relation between the momentum density and local velocity or current, is obtained by using
$e \jv=\sigmae\Ev+\sigma_{\rm ah}(\hat{\Mv}\times\Ev)$,
where $\sigma_{\rm e}$ and $\sigma_{\rm ah}$ are the longitudinal and anomalous Hall conductivities, respectively.
At the lowest order in the spin-orbit interaction,
the hydrodynamic equation of the present system reads
\begin{align}
 \dot{\pv}
 &=\tilde{\eta}_0 (2\nabla(\nabla\cdot\jv)+\nabla^2 \jv)
 + \tilde{\lambda} (2 \nabla (\omegav \cdot \hat{\Mv}) +\nabla^2(\jv \times \hat{\Mv}))
 + \tilde{f}_{\rm ah}(\jv \times \hat{\Mv})
 -\frac{n e^2}{\sigmae}\jv +en\Ev
 \label{pdot2}
\end{align}
where $\omegav\equiv \nabla\times\jv$ is the vortex density, $\tilde{\lambda}= e^2 \lambda/\sigmae$, $\tilde{\eta}_0\equiv e^2{\eta}_0/\sigmae$ and
$\tilde{f}_{\rm ah}\equiv e^2 f_{\rm ah}/\sigmae$.
The term $\tilde{\lambda} \nabla^2(\jv \times \hat{\Mv})$ is the anomalous Hall viscosity force arising from a non-dissipative component of the viscosity tensor when the fluids time-reversal symmetry is broken, which has been intensely discussed systems under a magnetic field~\cite{Avron98, Read09, Hoyos12, Scaffidi17, Berdyugin19}.
On the other hand, the term $\tilde{\lambda} \nabla (\omegav \cdot \hat{\Mv})$ is the pressure induced by the vorticity-magnetization coupling, which can also be regarded as an anomalous Hall contribution to the volume viscosity.
%%%%%%%%%%%%%%%%%%%%%%%%%%%%%%%%%%%%%%%%%%%%%%%%%%%%%%%
\begin{figure}
 \includegraphics[width=0.6\hsize]{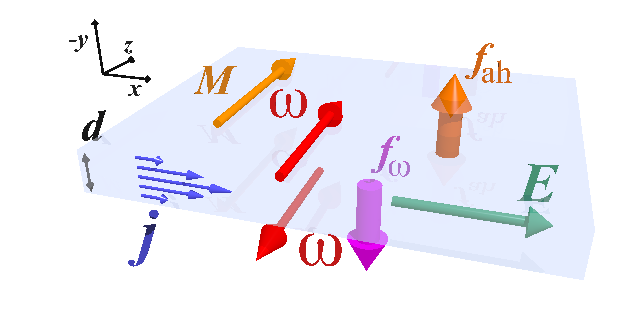}
 \caption{
 Schematic picture of the vorticity-induced anomalous Hall effect in a thin film with thickness $d$ with an in-plane magnetization (chosen along the $z$ axis).
Applied electric field $E$ along the $x$-direction induces an inhomogeneous fluid velocity $\vv$ in the thickness direction, resulting in positive and negative vorticities $\omegav$ close to the upper and lower plane, respectively.
 The gradient of the vorticity induces a Hall motive force $f_{\omega}$ perpendicular to the film, which is in the opposite direction of the conventional anomalous Hall force $f_{\rm ah}$.
 \label{FIGahe}}
\end{figure}
%%%%%%%%%%%%%%%%%%%%%%%%%%%%%%%%%%%%%%%%%%%%%%%%%%%%%%%%

%%%%%%%%%%%%%%%%%%%%%%%%%%%%%%%%%%%%%%%%%%%%%%%%%%%%%%%%
\section{Vorticity-induced anomalous Hall effect}
%%%%%%%%%%%%%%%%%%%%%%%%%%%%%%%%%%%%%%%%%%%%%%%%%%%%%%%%

Vorticity of flow is usually neglected in bulk transport phenomena in solids as the effect affects only near surfaces or interfaces.
In thin film or wires in contrast, the effect would dominate hydrodynamic transport.
In general theory of fluid, the fluid velocity at the interfaces with the container of fluid vanishes and the velocity grows away from the interface.
The velocity gradient means the existence of vorticity, $\omegav=\nabla\times\jv$, near the surfaces and interfaces.
Our result, Eq. (\ref{pdot2}), indicates that such surface vortices when couples to magnetization or magnetic field  induces a Hall force or voltage.
A suitable setting for observing this effect is a thin film ferromagnet with an in-plane magnetization (Fig. \ref{FIGahe}).
We apply an electric field $\Ev$ perpendicular to the magnetization.
In the steady state, the velocity profile in the thickness direction (shown in Fig. \ref{FIGahe}) calculated from Eq. (\ref{pdot2}) is
\begin{align}
j_x(y) = \frac{\sigmae}{e} E \left[1-\frac{\cosh(y/l)}{\cosh(d/2l)}\right],
\end{align}
where the length scale of the profile $l$ is determined by the viscosity and the electron mean free path as
$l\equiv \sqrt{\sigmae \tilde{\eta}_0/(n e^2)}$.
The velocity variation results in a vorticity, $\omegav=\nabla\times\jv$, positive near the upper plane and negative in the lower plane.
The vorticity-magnetization coupling energy $\omegav\cdot\hat{\Mv}$ has a gradient in the perpendicular direction, resulting in a vorticity-induced motive force density $\fv_{\omega}\equiv \tilde{\lambda}\nabla(\omegav\cdot\hat{\Mv})$.
Taking account of the elastic life time $\taue$, the steady perpendicular current density induced by the vorticity is
$j_{\omega}=f_{\omega}\taue/m$.
The vorticity due to the current density $j_x$ along the electric field is approximated as
$\omega=|\nabla\times\jv|\simeq j_x/l$ near the surface considering a system thickness $d\gtrsim l$.
The gradient of the vorticity is of the order of $\omega/d=j_x/(dl)$.
The vorticity-induced current density is therefore
$j_\omega=\sigma_\omega E$, where $\sigma_\omega\equiv \frac{\lambda\taue}{mdl}$ is the vorticity-induced anomalous Hall conductivity.
In the present model, the order of magnitudes of the coefficient $\lambda$ and the anomalous Hall conductivity are related by $\lambda\simeq(\ef\taue)\sigma_{\rm ah}$, and thus
\begin{align}
\sigma_\omega \simeq \sigma_{\rm ah}\frac{l}{d}
\end{align}
As the present hydrodynamic approach is justified in the regime $d\gtrsim l$, the vorticity-induced Hall effect is at most the same order of magnitude as the conventional anomalous Hall effect for a thin film of $d\sim l$.
Nevertheless, the present vorticity mechanism is a different origin of the Hall effect and could be useful for surface sensitive detection of transport.

\begin{figure}
 \includegraphics[width=0.45\hsize]{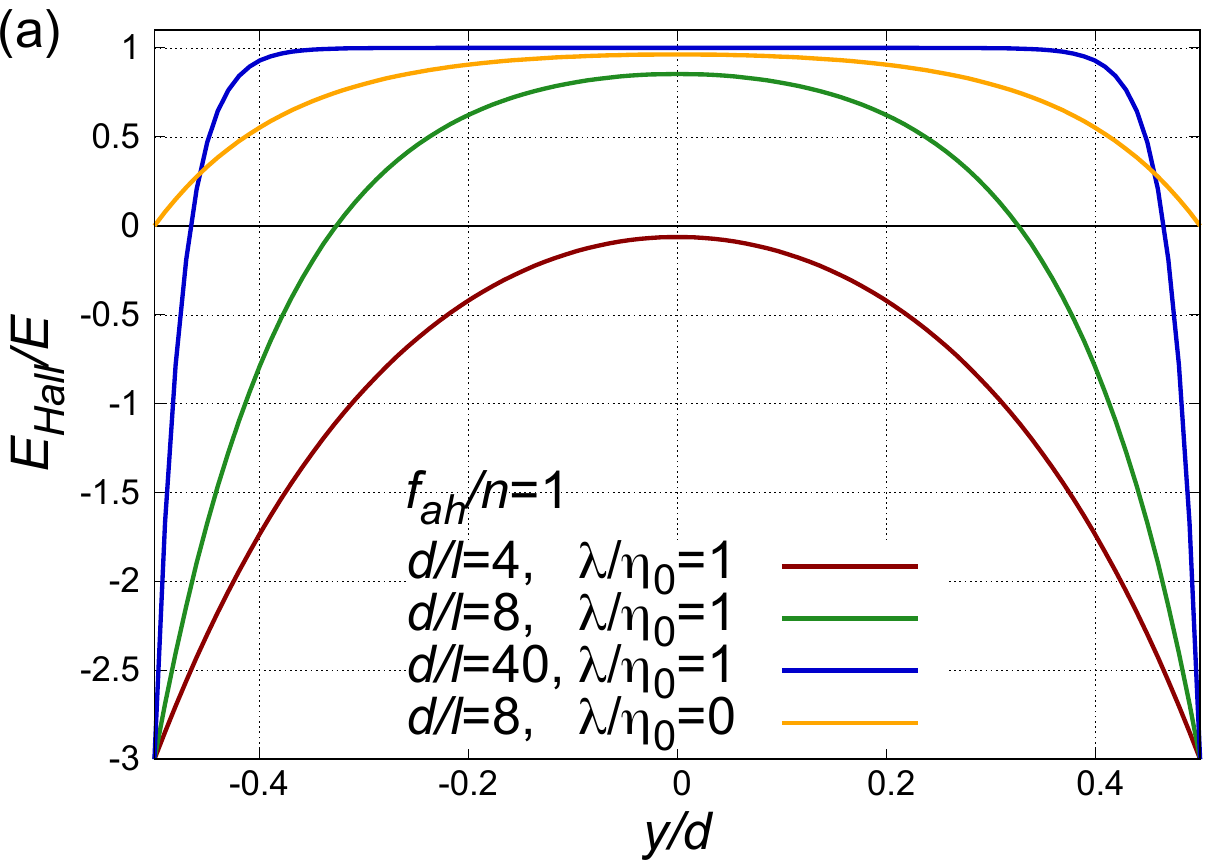}
 \includegraphics[width=0.45\hsize]{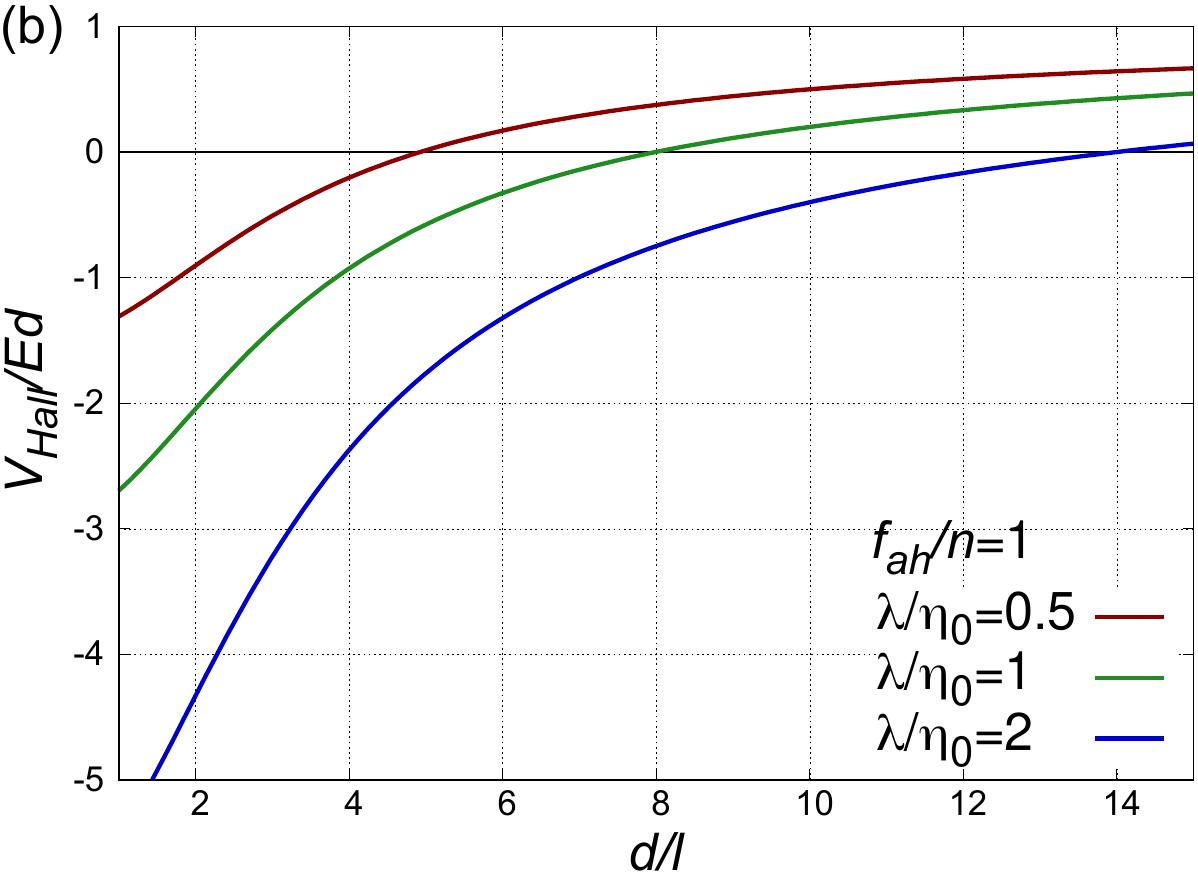}
 \caption{
 (a) Spatial distribution of Hall electric field (Eq. (\ref{E_Hall})). (b) System thickness dependence of the total Hall voltage (Eq. (\ref{V_Hall})), plotted for different value for $\lambda/\eta_0$.
 \label{FIGEVHall}}
\end{figure}

The profile of the total Hall electric field calculated from Eq. (\ref{pdot2}) is
\begin{align}
%E_{\rm Hall}(y) = -  \left[
%\frac {f_{\rm ah}}{n}  + \left( \frac{\tilde{\lambda}-\eta_{\rm ah}}{\tilde{\eta}_0}  - \frac {f_{\rm ah}}{n}
%\right)  \frac{\cosh(y/l)}{\cosh(d/2l)}
%\right]E,
E_{\rm Hall}(y) = \left[
\frac {f_{\rm ah}}{n}
 - \left( 3 \frac{ {\lambda} }{{\eta}_0}  + \frac {f_{\rm ah}}{n}
\right)  \frac{\cosh(y/l)}{\cosh(d/2l)}
\right]E,
\label{E_Hall}
\end{align}
where we included Hall viscosity for generality.
The Hall voltage is
\begin{align}
V_{\rm Hall} = \int_{-d/2}^{d/2}E_{\rm Hall}(y)dy
= \left[
\frac {f_{\rm ah}}{n} - \frac{2l}{d} \left( 3 \frac{\lambda}{\eta_0} + \frac {f_{\rm ah}}{n}
\right)  \tanh(d/2l)
\right]E d.
\label{V_Hall}
\end{align}
Here, the first term in the bracket is the conventional one, which is dominant in the bulk limit ($d\to \infty$), whereas the others are the corrections due to the vorticity or velocity gradient, which become important in mesoscopic systems ($d\sim l$).
Equation (\ref{E_Hall}) indicates that the bulk and vorticity contributions of the anomalous Hall effect have opposite signs, resulting in a negative Hall electric field near the boundaries as shown in Fig. \ref{FIGEVHall} (a).
In the bulk limit, $d\to \infty$, the total Hall voltage (Eq. (\ref{V_Hall})) reduces to the conventional contribution $\frac{f_{\rm ah}}{n}Ed$, while vorticity-induced negative contribution dominates in thin systems
with $d \sim l$ resulting in $V_{\rm Hall}\simeq -3\frac{\lambda}{\eta_0}Ed$ (Fig. \ref{FIGEVHall} (b)).
The sign change of the Hall voltage by changing the thickness would be useful for experimental identification of the vorticity-induced anomalous Hall effect.
The magnitudes of the two contributions are of the same order in our model;
$\lambda/\eta_0 \simeq f_{\rm ah}/n \simeq (\varepsilon_{\rm so}/\epsilon_{\rm F})({M}/{\epsilon_{\rm F}})$, where $\varepsilon_{\rm so}\equiv  \lambda_{\rm so} \vimp k_{\rm F}^2 $ is the energy scale of the spin-orbit interaction, while they would behave differently in general.

%%%%%%%%%%%%%%%%%%%%%%%%%%%%%%%%%%%%%%%%%%%%%%%%%%%%%%%
%%%%%%%%%%%%%%%%%%%%%%%%%%%%%%%%%%%%%%%%%%%%%%%%%%%%%%%%

\section{Conclusion}

We have derived a hydrodynamic equation for anomalous Hall electron fluid in the ohmic regime, and found a vorticity-magnetization coupling contribution in the momentum flux density.
The coupling induces a vorticity-induced additional anomalous Hall effect
in the opposite direction as the conventional anomalous Hall effect,
suggesting a sign reversal of anomalous Hall voltage in thin systems.

For vorticity to arise, existence of a boundary is generally essential, as the vorticity is parity invariant, while linear driving field is odd. A inversion symmetry breaking due to the boundary is thus necessary.

Spin chirality $\chiv$ has been pointed out to be an another origin of the anomalous Hall effect.
The study of the perturbative regime in Ref. \cite{TK02} is straightforwardly extended to the calculation of the momentum flux density, resulting in a coupling $\chiv\cdot\omegav$.

%%%%%%%%%%%%%%%%%%%%%%%%%%%%%%%%%%%%%%%%%%%%%%%%%%%%%%%%%%%
\acknowledgements
%The authors thank S. Seki and Y. Fuseya for valuable discussion.
This study was supported by
a Grant-in-Aid for Scientific Research (B) (No. 17H02929) from the Japan Society for the Promotion of Science and JSPS KAKENHI (Grant No. 20J22612).
\appendix

%%%%%%%%%%%%%%%%%%%%%%%%%%%%%%%%%%%%%%%
\section{Normal viscosity of ohmic fluid \label{SECnormalviscosity}}
%%%%%%%%%%%%%%%%%%%%%%%%%%%%%%%%%%%%%%%
Here we calculate the normal viscosity constant in the present ohmic fluid for a consistency.
The normal contribution to the momentum flux density is
\begin{align}
 \pi^{(0)}_{ij}(\rv,t) &\equiv -i\frac{1}{m} \tr[\hat{p}_i \hat{p}_j G^<(\rv,t,\rv,t)]
\end{align}
where the Green's function here is without the spin-orbit interaction.
Taking account of the static external electric field, described by a gauge field $\Av$ as $\Ev=-\dot{\Av}$,
to the linear order, it reads in the Fourier representation
$ \pi^{(0)}_{ij}(\qv,\Omega) =
 \int d^3r \int dt
 \pi^{(0)}_{ij}(\rv,t) e^{-i {\bm q} \cdot {\bm r}} e^{i \Omega t} $
\begin{align}
 \pi^{(0)}_{ij}(\qv) &= i\frac{e}{m^2}
 \frac{1}{V} \sum_{\kv}\sumom \lim_{\Omega\ra0}
  k_i k_j k_k [G_{\kv+\frac{\qv}{2},\omega+\Omega}
  G_{\kv-\frac{\qv}{2},\omega}]^< A_k(\qv,\Omega)
\nnr
  & \equiv
 \pi^{(0)}_{ijk}(\qv) eE_k(\qv)
\end{align}
where $\omega$ and $\kv$, $\Omega$ and $\qv$ denote the angular frequency and wave vector of electron and external field, respectively.
Here we neglect the diffusive contribution as it affects only the non equilibrium state.
The response function is
\begin{align}
 \pi^{(0)}_{ijk}(\qv) &\equiv \frac{1}{m^2}
\ frac{1}{V} \sum_{\kv}\sumom \lim_{\Omega\ra0}\frac{1}{\Omega}
  k_i k_j k_k [G_{\kv+\frac{\qv}{2},\omega+\Omega}
  G_{\kv-\frac{\qv}{2},\omega}]^<
\end{align}
The lesser component is decomposed into the retarded and advanced Green's functions  as
\begin{align}
[G_{\kv+\frac{\qv}{2},\omega+\Omega}G_{\kv-\frac{\qv}{2},\omega}]^<
&= (f_{\omega} -f_{\omega+\Omega}) G_{\kv+\frac{\qv}{2},\omega+\Omega}^\ret G_{\kv-\frac{\qv}{2},\omega}^\adv
- f_{\omega} G_{\kv+\frac{\qv}{2},\omega+\Omega}^\ret G_{\kv-\frac{\qv}{2},\omega}^\ret
+ f_{\omega+\Omega} G_{\kv+\frac{\qv}{2},\omega+\Omega}^\adv G_{\kv-\frac{\qv}{2},\omega}^\adv
\end{align}
where $f_\omega=[e^{\beta\omega}+1]^{-1}$ is the Fermi distribution function.
Expanding the expression with respect to $\Omega$, we obtain
\begin{align}
[G_{\kv+\frac{\qv}{2},\omega+\Omega} G_{\kv-\frac{\qv}{2},\omega}]^<
&=
 f_{\omega} [G_{\kv+\frac{\qv}{2},\omega}^\adv G_{\kv-\frac{\qv}{2},\omega}^\adv
 -G_{\kv+\frac{\qv}{2},\omega}^\ret G_{\kv-\frac{\qv}{2},\omega}^\ret
]\nnr
&+
\Omega f'_{\omega}\lt[
-G_{\kv+\frac{\qv}{2},\omega}^\ret G_{\kv-\frac{\qv}{2},\omega}^\adv
+ \frac{1}{2} (
G_{\kv+\frac{\qv}{2},\omega}^\adv G_{\kv-\frac{\qv}{2},\omega}^\adv
+
G_{\kv+\frac{\qv}{2},\omega}^\ret G_{\kv-\frac{\qv}{2},\omega}^\ret ) \rt]
\nnr &
+\frac{\Omega}{2}f_\omega [
-G_{\kv+\frac{\qv}{2},\omega}^\adv\stackrel{\leftrightarrow}{\partial_\omega} G_{\kv-\frac{\qv}{2},\omega}^\adv
+G_{\kv+\frac{\qv}{2},\omega}^\ret \stackrel{\leftrightarrow}{\partial_\omega} G_{\kv-\frac{\qv}{2},\omega}^\ret] \label{GGless}
\end{align}
The contribution of the order of $\Omega^0$ turns out to cancel the contribution from the anomalous velocity $\delta v$.
The last term of the right-hand side of Eq. (\ref{GGless}) is smaller than the main contribution, the second term, by a factor of
$(\ef\taue)^{-1}$ and is neglected.
The response function is therefore, using $f'(\omega)=-\delta(\omega)$ at low temperatures,
\begin{align}
 \pi^{(0)}_{ijk}(\qv) &\equiv \frac{1}{m^2}
 \frac{1}{V} \sum_{\kv}\sumom f'(\omega)  k_i k_j k_k
  \lt[
-G_{\kv+\frac{\qv}{2},\omega}^\ret G_{\kv-\frac{\qv}{2},\omega}^\adv
+ \frac{1}{2} (
G_{\kv+\frac{\qv}{2},\omega}^\adv G_{\kv-\frac{\qv}{2},\omega}^\adv
+
G_{\kv+\frac{\qv}{2},\omega}^\ret G_{\kv-\frac{\qv}{2},\omega}^\ret ) \rt]
\nnr
& = - \frac{1}{2\pi}\frac{1}{m^2}
 \frac{1}{V} \sum_{\kv} k_i k_j k_k
  \lt[
-G_{\kv+\frac{\qv}{2},\omega}^\ret G_{\kv-\frac{\qv}{2},\omega}^\adv
+ \frac{1}{2} ( G_{\kv+\frac{\qv}{2},\omega}^\adv G_{\kv-\frac{\qv}{2},\omega}^\adv
+ G_{\kv+\frac{\qv}{2},\omega}^\ret G_{\kv-\frac{\qv}{2},\omega}^\ret ) \rt]
\end{align}
where $G_{\kv}\equiv G_{\kv,\omega=0}$.
Expanding with respect to $q$ and using rotational symmetry in $\kv$, we obtain
\begin{align}
 \pi^{(0)}_{ijk}(\qv)
& =  \frac{- 1}{4\pi V}\frac{iq_l}{m^3 \taue} \sum_{\kv} k_i k_j k_k k_l
|G_{\kv}^\ret|^4 \nnr
& = - \eta_0 i[\delta_{ij}q_k+\delta_{ik}q_j+\delta_{jk}q_i]
\end{align}
where $\eta_0\equiv \frac{\nu\kf^4\taue^2}{15m^3}$ is a normal viscosity constant.
Using $\nu\sim\ef$, it is $\eta_0\simeq l_{{\rm e}}^2$, where $l_{{\rm e}}\equiv \frac{\kf}{m}\taue$ is the elastic mean free path.
Including the external field,
\begin{align}
 \pi^{(0)}_{ij}
& = - e\eta_0 [\delta_{ij}\nabla\cdot\Ev+\nabla_{i}E_j+\nabla_{j}E_i]
\end{align}
and the contribution to the time-derivative of momentum density is
\begin{align}
 - \nabla_j \pi^{(0)}_{ij}
& = e\eta_0 [2 \nabla(\nabla\cdot\Ev)+\nabla^2 \Ev]
\end{align}

%%%%%%%%%%%%%%
%\bibliography{/home/gt/References/15,/home/gt/References/gt,/home/gt/References/remarks}
%\bibliography{/home/funaki/c_home/bibtex/from-prof/15}

\end{document}